\documentclass[12pt,psfig]{article}

\usepackage{psfig}

\usepackage{scicite}

\usepackage{times}

\newenvironment{sciabstract}{%
\begin{quote} \bf}
{\end{quote}}

\newcounter{lastnote}
\newenvironment{scilastnote}{%
\setcounter{lastnote}{\value{enumiv}}%
\addtocounter{lastnote}{+1}%
\begin{list}%
{\arabic{lastnote}.}
{\setlength{\leftmargin}{.22in}}
{\setlength{\labelsep}{.5em}}}
{\end{list}}

\title{Tests of general relativity from timing the double pulsar}

\author{M. Kramer,$^{1\ast}$ I.H. Stairs,$^{2}$ R.N. Manchester,$^{3}$
  M.A. McLaughlin,$^{1,4}$ \\
A.G. Lyne,$^{1}$ R.D. Ferdman,$^{2}$ M. Burgay,$^{5}$ D.R. Lorimer,$^{1,4}$ \\
 A. Possenti,$^{5}$  N. D'Amico,$^{5,6}$ 
  J.M. Sarkissian,$^{3}$ G.B. Hobbs,$^{3}$ \\
J.E. Reynolds,$^{3}$ P.C.C. Freire$^{7}$ and F. Camilo$^{8}$ \\
\\
\normalsize{$^{1}$University of Manchester, Jodrell Bank Observatory, Macclesfield, SK11 9DL, UK}\\
\normalsize{$^{2}$Dept. of Physics and Astronomy, University of British
Columbia, 6224 Agricultural Road,}\\
\normalsize{Vancouver, BC V6T 1Z1, Canada}\\
\normalsize{$^{3}$Australia Telescope National Facility, CSIRO, P.O.~Box~76, Epping
NSW~1710, Australia}\\
\normalsize{$^{4}$Department of Physics, West Virginia University, Morgantown,
WV 26505, USA}\\
\normalsize{$^{5}$INAF - Osservatorio Astronomica di Cagliari, Loc. Poggio dei Pini, Strada 54,}\\
\normalsize{09012 Capoterra, Italy}\\
\normalsize{$^{6}$Universita' degli Studi di Cagliari, Dipartimento di Fisica, SP
   Monserrato-Sestu km 0.7,} \\
\normalsize{ 09042 Monserrato (CA), Italy}\\
\normalsize{$^{7}$NAIC, Arecibo Observatory, HC03 Box 53995, PR 00612, USA}\\
\normalsize{$^{8}$Columbia Astrophysics Laboratory, Columbia University, 550 West 120$^{th}$ Street,}\\
\normalsize{New York, NY 10027, USA}\\
\\
\normalsize{$^\ast$To whom correspondence should be addressed; E-mail: mkramer@jb.man.ac.uk}
}

\date{}

\begin{document} 


\baselineskip24pt


\maketitle


\begin{sciabstract}
  The double pulsar system, PSR J0737-3039A/B, is unique in that both neutron
  stars are detectable as radio pulsars. This, combined with significantly
  higher mean orbital velocities and accelerations when compared to other
  binary pulsars, suggested that the system would become the best available
  testbed for general relativity and alternative theories of gravity in the
  strong-field regime. Here we report on precision timing observations taken
  over the 2.5 years since its discovery and present four independent
  strong-field tests of general relativity.  Use of the theory-independent
  mass ratio of the two stars makes these tests uniquely different from
  earlier studies. By measuring relativistic corrections to the Keplerian
  discription of the orbital motion, we find that the ``post-Keplerian''
  parameter $s$ agrees with the value predicted by Einstein's theory of
  general relativity within an uncertainty of 0.05\%, the most precise test
  yet obtained.  We also show that the transverse velocity of the system's
  center of mass is extremely small.  Combined with the system's location near
  the Sun, this result suggests that future tests of gravitational theories
  with the double pulsar will supersede the best current Solar-system tests.
  It also implies that the second-born pulsar may have formed differently to
  the usually assumed core-collapse of a helium star.
\end{sciabstract}



\paragraph*{Introduction.}

Einstein's general theory of relativity (GR) has so far passed all
experimental tests with flying colours\cite{wil01}, with the most
precise tests achieved in the weak-field gravity conditions of the Solar
System\cite{bit03,wtb04}.  However, it is conceivable that GR breaks
down under extreme conditions such as strong gravitational fields
where other theories of gravity may apply\cite{de98}. 
Predictions of gravitational radiation and
self-gravitational effects can only be tested using massive and
compact astronomical objects such as neutron stars and black
holes. Studies of the double-neutron-star binary systems, PSR B1913+16
and PSR B1534+12, have provided the best such tests so
far, confirming GR at the 0.2\% and 0.7\% level,
respectively\cite{tw89,sttw02}
\footnote{
Stairs et al.~(2002, ref.~\cite{sttw02}) 
find an agreement of their measured
values for PSR~B1534+12 with GR at the 0.05\% level, but the measurement
uncertainty on the most precisely measured parameter in the test, $s$, is
only 0.7\%.}.
The recently discovered double pulsar system,
PSR J0737-3039A/B, has significantly higher mean orbital velocities
and accelerations than either PSR B1913+16 or PSR B1534+12 and is
unique in that both neutron stars are detectable as radio
pulsars\cite{bdp+03,lbk+04}.

PSR J0737$-$3037A/B consists of a 22-ms period pulsar, PSR J0737$-$3039A
(henceforth called A), in a 2.4-hr orbit with a younger 2.7-s period pulsar,
PSR J0737$-$3039B (B).  Soon after the discovery of A\cite{bdp+03}, it was
recognised that the orbit's orientation, measured as the longitude of
periastron $\omega$, was changing in tine with a very large rate of
$\dot{\omega} =d\omega/dt \sim 17^\circ$ yr$^{-1}$, which is four times the
corresponding value for the Hulse-Taylor binary, PSR B1913+16\cite{tw89}. This
immediately suggested that the system consists of two neutron stars, a
conclusion confirmed by the discovery of pulsations from B\cite{lbk+04}. The
pulsed radio emission from B has a strong orbital modulation, both in
intensity and in pulse shape. It appears as a strong radio source only for two
intervals, each of about 10-min duration, while its pulsed emission is rather
weak or even undetectable for most of the remainder of the
orbit\cite{lbk+04,bpm+05}.

In double-neutron-star systems, especially those having short orbital periods,
observed pulse arrival times are significantly modified by relativistic
effects which can be modelled in a theory-independent way using the so-called
``Post-Keplerian'' (PK) parameters\cite{dd86}.  These PK parameters are
phenomenological corrections and additions to the simple Keplerian description
of the binary motion, describing for instance a temporal change in period or
orientation of the orbit, or an additional ``Shapiro-delay'' that occurs due
to the curvature of space-time when pulses pass near the massive companion.
The PK parameters take different forms in different theories of gravity and so
their measurement can be used to test these theories\cite{dt92,wil01}. For
point masses with negligible spin contributions, GR predicts values for the PK
parameters which depend only on the two a priori unknown neutron-star masses
and the precisely measurable Keplerian parameters. Therefore measurement of
three (or more) PK parameters provides one (or more) tests of the predictive
power of GR. For the double pulsar we can also measure the mass ratio of the
two stars, $R \equiv m_{\rm{A}}/m_{\rm{B}} = x_{\rm{B}}/x_{\rm{A}}$. The
ability to measure this quantity provides an important constraint because in
GR and other theories this simple relationship between the masses and
semi-major axes is valid to at least first post-Newtonian (1PN) or $(v/c)^2$
order\cite{ds88,dt92}.

\paragraph*{Observations.}

Timing observations of PSR J0737$-$3039A/B have been undertaken using the 64-m
Parkes radio telescope in New South Wales, Australia, 
the 76-m Lovell radio telescope at
Jodrell Bank Observatory (JBO), UK, and the 100-m Green Bank Telescope (GBT) 
in West Virginia, USA, between 2003 April and 2006 January. 

At Parkes, observations were carried out in bands centred 
at 680 MHz, 1374 MHz and 3030 MHz.
While timing observations were frequent after the discovery of the system,
later observations at Parkes were typically conducted every 3-4 weeks, usually
covering two full orbits per session. Observations at the GBT were conducted
at monthly intervals, with each session consisting of a 5- to 8-hour track
(i.e., 2 to 3 orbits of the double pulsar). Typically, the observing
frequencies were 820 and 1400 MHz for alternate sessions.  Occasionally, we
also performed observations at 340\,MHz, in conjunction with pulse profile
studies to be reported elsewhere.  In addition, we conducted concentrated
campaigns of five 8-hour observing sessions, all at 820 MHz, in 2005 May and
2005 November. Observations at JBO employed the 76-m Lovell telescope.  Most
data were recorded at 1396 MHz, while some observing sessions were carried out
at the lower frequency of 610 MHz. The timing data obtained at Jodrell Bank
represent the most densely sampled dataset but, because of the limited
bandwidth, requiring longer integration times per timing point.
The Parkes dataset is the
longest one available and hence provides an excellent basis for investigation
of secular timing terms.  

The time-series data of all systems were folded modulo the predicted
topocentric pulse period.  The adopted integration times were 30~s for pulsar
A (180~s for JBO data) and 300~s for pulsar B. For A, these integration times
reflect a compromise between producing pulse profiles with adequate
signal-to-noise ratio and sufficient sampling of the orbit to detect and
resolve phenomena that depend on orbital phase, such as the Shapiro delay. The
integration time for B corresponds to about 108 pulse periods and is a
compromise between the need to form a stable pulse profile while resolving the
systematic changes seen as a function of orbital phase.

\paragraph*{Timing measurements.}

For each of the final profiles, pulse times-of-arrival (TOAs) were computed by
correlating the observed pulse profiles with synthetic noise-free templates
(see Fig.~1 in \cite{som},
 cf.~ref.~\cite{kll+99}). A total of 131,416 pulse TOAs were
measured for A while 507 TOAs were obtained for B. For A, the same template
was used for all observations in a given frequency band, but different
templates were used for widely separated bands.  We note that our observations
still provide no good evidence for secular evolution of A's
profile\cite{mkp+05} despite the predictions of geodetic precession.  The best
timing precision was obtained at 820 MHz with GASP backend (see 
ref.~\cite{som} for details of this and other observing systems) on the
GBT, with typical TOA measurement uncertainties for pulsar A of 18 $\mu$s for
a 30-s integration.

For B, because of the orbital and secular dependence of its pulse
profile\cite{bpm+05}, different templates were also used for different orbital
phases and different epochs. A matrix of B templates was constructed, dividing
the data set into 3-month intervals in epoch and 5-minute intervals in orbital
phase. The results for the 29 orbital phase bins were studied, and it was
noticed that, while the profile changes dramatically and quickly during the
two prominent bright phases, the profile shape is simpler and more stable at
orbital phases when the pulsar is weak. This apparent stability at some
orbital phases cannot be attributed to a low signal-to-noise ratio as secular
variations in the pulse shape are still evident.  Consequently, the orbital
phase was divided into five groups of different lengths to which the
same template (for a given 3-month interval) was applied as shown in Fig.~2
of \cite{som}.
In the final timing analysis, data from the two groups representing the bright
phases (IV \& V in Fig.~2 of \cite{som}) 
were excluded to minimize the systematic errors
caused by the orbital profile changes. Also, because of signal-to-noise and
radio interference considerations, only data from Parkes and the GBT BCPM
backend were used in the B timing analysis.

All TOAs were transferred to Universal Coordinated Time (UTC) using the Global
Positional System (GPS) to measure offsets of station clocks from national
standards and Circular T of the BIPM to give offsets from UTC, and then to the
nominally uniform Terrestrial Time TT(BIPM) timescale. These final TOAs were
analysed using the standard software package {\sc tempo} \cite{tempo}, 
fitting parameters
according to the relativistic and theory independent timing model of Damour \&
Deruelle\cite{dd85,dd86}.  In addition to the DD model, we also applied the
``DD-Shapiro'' (DDS) model introduced by Kramer et al.~(ref.~\cite{ksm+06a}).
The DDS model is a modification of the DD model designed for highly inclined
orbits. Rather than fitting for the Shapiro parameter $s$, the model uses the
parameter $z_s \equiv -\ln(1-s)$ which gives a more reliable determination of
the uncertainties in $z_s$ and hence in $s$.  We quote the final result for
the more commonly used parameter $s$ and note that its value computed from
$z_s$ is in good agreement with the value obtained from a direct fit for $s$
within the DD model. Derived pulsar and binary system parameters are listed in
Table~1.

In the timing analysis for pulsar B, we used an unweighted fit to avoid
biasing the fit toward bright orbital phases. Uncertainties in the timing
parameters were estimated using Monte Carlo simulations of fake data sets for
a range of TOA uncertainties, ranging from the minimum estimated TOA error to
its maximum observed value of about 4 ms. For B, we also fitted for offsets
between datasets derived from different templates in the fit since the
observed profile changes prevent the establishment of a reliable phase
relationship between the derived templates.  This precludes a coherent fit
across the whole orbit and hence limits the final timing precision for B.  It
cannot yet be excluded that different parts of B's magnetosphere are active
and responsible for the observed emission at different orbital phases.

In the final fit, we adopted the astrometric parameters and the dispersion
measure derived for A and held these fixed during the fit, since A's shorter
period and more stable profile give much better timing precision than is
achievable for B.  Except for the semi-major axis which is only observable as
the projection onto the plane-of-the-sky $x_{\rm{B}}=(a_{\rm{B}}/c)\sin i$,
where $i$ is the orbital inclination angle, we also adopted A's Keplerian
parameters (with $180^\circ$ added to $\omega_{\rm{A}}$) and kept these fixed.
We also adopted the PK parameter $\dot{\omega}$ from the A fit since logically
this must be identical for the two pulsars; this equality therefore does not
implicitly make assumptions about the validity of any particular theory of
gravity (see next section). The same applies for $\dot{P}_{\rm b}$. In
contrast, the PK parameters $\gamma$, $s$ and $r$ are asymmetric in the masses
and their values and interpretations differ for A and B.  In practical terms,
the relatively low timing precision for B does not require the inclusion of
$\gamma$, $s$, $r$ or $\dot{P}_{\rm b}$ in the timing model. We can however
independently measure $\dot\omega_{\rm B}$, obtaining a value of
$16.96\pm0.05$ deg yr$^{-1}$, consistent with the more accurately determined
value for A.

Since the overall precision of our tests of GR is currently limited by our
ability to measure $x_{\rm{B}}$ and hence the mass ratio $R \equiv
m_{\rm{A}}/m_{\rm{B}} = x_{\rm{B}}/x_{\rm{A}}$ (see below), we adopted the
following strategy to obtain the best possible accuracy for this parameter. We
used the whole TOA data set for B in order to measure B's spin parameters $P$
and $\dot{P}$, given in Table 1. These parameters were then
kept fixed for a separate analysis of the concentrated 5-day GBT observing
sessions at 820 MHz.  On the timescale of the long-term profile evolution of
B, each 5-day session represents a single-epoch experiment and hence requires
only a single set of profile templates. The value of $x_{\rm{B}}$ obtained
from a fit of this parameter only to the two 5-day sessions is presented in
Table 1.

Because of the possible presence of unmodelled intrinsic pulsar timing noise
and because not all TOA uncertainties are well understood, we adopt the common
and {\em conservative} pulsar-timing practice of reporting twice the parameter
uncertainties given by {\sc tempo} as estimates of the 1-$\sigma$
uncertainties.  While we believe that our real measurement uncertainties are
actually somewhat smaller than quoted, this practice facilitates the
comparison with previous tests of GR using pulsars.  The timing model also
includes timing offsets between the datasets for the different instruments
represented by the entries in Table~1 in \cite{som}.  The final weighted rms
post-fit residual is $54.2\mu$s.  In addition to the spin and astrometric
parameters, the Keplerian parameters of A's orbit and five PK parameters, we
also quote a tentative detection of a timing annual parallax which is
consistent with the dispersion-derived distance. Further details are given in
ref.~\cite{som}.

\paragraph*{Tests of general relativity.}

Previous observations of PSR J0737$-$3039A/B\cite{bdp+03,lbk+04} resulted in
the measurement of $R$ and four PK parameters: the rate of periastron advance
$\dot{\omega}$, the gravitational redshift and time dilation parameter
$\gamma$, and the Shapiro-delay parameters $r$ and $s$.  Compared to these
earlier results, the measurement precision for these parameters from PSR
J0737$-$3039A/B has increased by up to two orders of magnitude. Also, we have
now measured the orbital decay, $\dot{P}_{\rm b}$. Its value, measured at the
1.4\% level after only 2.5 years of timing, corresponds to a shrinkage of the
pulsars' separation at a rate of 7mm per day. Therefore, we have measured
five PK parameters for the system in total. 
Together with the mass ratio $R$, we have six
different relationships that connect the two unknown masses for A and B with
the observations. Solving for the two masses using $R$ and a one PK parameter,
we can then use each further PK parameter to compare its observed value with
that predicted by GR for the given two masses, providing four independent
tests of GR. Equivalently, one can display these tests elegantly in a
``mass-mass'' diagram (Fig.~1).  Measurement of the PK parameters gives curves
on this diagram that are in general different for different theories of
gravity but which should intersect in a single point, i.e., at a pair of mass
values, if the theory is valid\cite{dt92}.

As shown in Fig.~1,
we find that all measured constraints are consistent with GR. The
most precisely measured PK parameter currently available is the precession of
the longitude of periastron, $\dot{\omega}$. We can combine this with the
theory-independent mass ratio $R$ to derive the masses given by the
intersection region of their curves: $m_{\rm{A}} = 1.3381\pm0.0007$ M$_\odot$
and $m_{\rm{B}} = 1.2489\pm0.0007$ M$_\odot$.\footnote{The true masses will
  deviate from these values by an unknown, but essentially constant, Doppler
  factor, probably of order $10^{-3}$ or less\cite{dd86}. Moreover,
  what is measured is a product containing Newton's gravitational constant
  $G$. The relative uncertainty of $G$ of $1.5 \times 10^{-4}$ limits our
  knowledge of any astronomical mass in kilograms but since the product 
  $T_\odot = GM_{\odot}/c^3 = 4.925490947\mu$s 
  is known to very high precision, masses can be measured precisely
  in solar units.}
Table~2 lists the
resulting four independent tests that are currently available.  All of them
rely on comparison of our measured values of $s$, $r$, $\gamma$ and $\dot
P_{\rm b}$ with predicted values based on the masses defined by the
intersection of the allowed regions for $\dot\omega$ and $R$ in the
$m_A$--$m_B$ plane. The calculation of the predicted values is somewhat
complicated by the fact that the orbit is nearly edge-on to the line of sight,
so that the formal intersection region actually includes parts of the plane
disallowed by the Keplerian mass functions of both pulsars (see Fig.~1).
To derive legitimate predictions for the various parameters, we
used the following Monte Carlo method.  A pair of trial values for
$\dot\omega$ and $x_B$ (and hence $R$ and the B mass function) is selected
from gaussian distributions based on the measured central values and
uncertainties. (The uncertainty on $x_A$ is very small and is neglected in
this procedure.)  This pair of trial values is used to derive trial masses
$m_A$ and $m_B$, using the GR equation $\dot \omega = 3
(\frac{P_b}{2\pi})^{-5/3} (T_\odot M)^{2/3}\,(1-e^2)^{-1}$, where $M =
m_A+m_B$ and $T_\odot\equiv GM_\odot/c^3 = 4.925490947\,\mu$s, and the
mass-ratio equation $m_{\rm{A}}/m_{\rm{B}} = x_{\rm{B}}/x_{\rm{A}}$. If this
trial mass pair falls in either of the two disallowed regions (based on the
trial mass function for $B$) it is discarded. This procedure allows for the
substantial uncertainty in the B mass function. Allowed mass pairs are 
then used to
compute the other PK parameters, assuming GR.  This procedure is repeated
until large numbers of successful trials have accumulated.  Histograms of the
PK predictions are used to compute the expectation value and 68\% confidence
ranges for each of the parameters.  These are the values given in Table~2.

The
Shapiro delay shape illustrated in Fig.~2 gives the most precise test,
with $s_{\rm obs}/s_{\rm pred} = 0.99987\pm0.00050$.\footnote{
Note, $s$ has the same relative uncertainty as our determination of the
masses.}  This is by far
the best available test of GR in the strong-field limit, having a
higher precision than the test based on the observed orbit decay in
the PSR B1913+16 system with a 30-year data span\cite{wt05}. As for
the PSR~B1534+12 system\cite{sttw02}, the PSR J0737$-$3039A/B
Shapiro-delay test is complementary to that of B1913+16 since it is
not based on predictions relating to emission of gravitational
radiation from the system\cite{twdw92}. Most importantly, the four
tests of GR presented here are qualitatively different
from all previous tests because they include one constraint ($R$) that
is independent of the assumed theory of gravity at the 1PN order. As
a result, for any theory of gravity, the intersection point is expected
to lie on the mass ratio line in Fig.~1. GR also passes this 
additional constraint.

In estimating the final uncertainty of $x_{\rm{B}}$ and hence of $R$, we have
considered that geodetic precession will lead to changes to the system
geometry and hence changes to the aberration of the rotating pulsar beam.  The
effects of aberration on pulsar timing are usually not separately measurable
but are absorbed into a redefinition of the Keplerian parameters. As a result,
the {\em observed} projected sizes of the semi-major axes, $x^{\rm obs}_{\rm
  A, B}$, differ from the {\em intrinsic sizes}, $x^{\rm int}_{\rm A, B}$ by a
factor $(1+\epsilon^{\rm A}_{\rm A, B})$. The quantity $\epsilon_{\rm A}$
depends for each pulsar A and B on the orbital period, the spin frequency,
the orientation of the pulsar spin and the
system geometry\cite{dt92}. While aberration should
eventually become detectable in the timing, allowing the determination of a
further PK parameter, at present it leads to an undetermined deviation of
$x^{\rm obs}$ from $x^{\rm int}$, where the latter is the relevant quantity
for the mass ratio. The parameter $\epsilon^{\rm A}_{\rm A, B}$ scales with
pulse period and is therefore expected to be two orders of magnitude smaller
for A than for B. However, because of the high precision of the A timing
parameters, the derived value $x^{\rm obs}_{\rm A}$ may already be
significantly affected by aberration. This has (as yet) no consequences for
the mass ratio $R=x^{\rm obs}_{\rm B}/x^{\rm obs}_{\rm A}$, as the uncertainty
in $R$ is dominated by the much less precise $x^{\rm obs}_{\rm B}$. We can
explore the likely aberration corrections to $x^{\rm obs}_{\rm B}$ for various
possible geometries. Using a range of values given by studies of the double
pulsar's emission properties\cite{lyu05}, we estimate $\epsilon^{\rm A}_{\rm
  A} \sim 10^{-6}$ and $\epsilon^{\rm A}_{\rm B} \sim 10^{-4}$. The
contribution of aberration therefore is at least one order of magnitude
smaller than our current timing precision.  In the future this effect may
become important, possibly limiting the usefulness of $R$ for tests of GR. If
the geometry cannot be independently determined, we could use the observed
deviations of $R$ from the value expected within GR to determine
$\epsilon^{\rm A}_{\rm B}$ and hence the geometry of B.

\paragraph*{Space motion and inclination of the orbit.}

Because the measured uncertainty in $\dot{P}_{\rm b}$ decreases approximately
as $T^{-2.5}$, where $T$ is the data span, we expect to improve our test of
the radiative aspect of the system to the 0.1\% level or better in about five
years' time.  For the PSR B1913+16 and PSR B1534+12 systems, the precision of
the GR test based on the orbit-decay rate is severely limited both by the
uncertainty in the differential acceleration of the Sun and the binary system
in the Galactic gravitational potential and the uncertainty in pulsar
distance\cite{dt91,sttw02}. For PSR J0737$-$3039A/B, both of these corrections
are very much smaller than for these other systems. 
Based on the measured dispersion measure and a model
for the Galactic electron distribution\cite{cl02}, PSR J0737$-$3039A/B is
estimated to be about 500 pc from the Earth. From the timing data we have
measured a marginally significant value for the annual parallax, $3\pm2$ mas,
corresponding to a distance of $200-1000$ pc (Table 1), 
which is consistent with the
dispersion-based distance that was also used for studies of detection rates in
gravitational wave detectors \cite{bdp+03}.  The observed proper motion of the
system (Table~1) and differential acceleration in the Galactic
potential\cite{kg89} then imply a kinematic correction to $\dot{P}_{\rm b}$ at
the 0.02\% level or less.  Independent distance estimates also can be expected
from measurements of the annual parallax by Very Long Baseline Interferometry
(VLBI) observations, allowing a secure compensation for this already small
effect. A measurement of $\dot{P}_{\rm b}$ at the 0.02\% level or better will
provide stringent tests for alternative theories of gravity.  For example,
limits on some scalar-tensor theories will surpass the best current
Solar-system tests\cite{de06}.

In GR, the parameter $s$ can be identified with $\sin i$ where $i$ is the
inclination angle of the orbit. The value of $s$ given in Table~1 corresponds
to $i= 88^\circ.69^{+0^\circ.50}_{-0^\circ.76}$.  Based on scintillation
observations of both pulsars over the short time interval when A is close to
superior conjunction, Coles et al.\cite{cmr+05} derived a value for
$|i-90^\circ|$ of $0^\circ.29 \pm 0^\circ.14$. This is consistent with our
measurement only at the 3-$\sigma$ level. As mentioned above, we used the DDS
model to solve for the Shapiro delay. Fig.~3 shows the resulting $\chi^2$
contours in the $z_s$ -- $m_{\rm B}$ plane. The value and uncertainty range
for $s$ quoted in Table 1 correspond to the peak and range of the 68\%
contour.  Because of the non-linear relationship between $z_s$ and $s$, the
uncertainty distribution in $s$ (and hence in $i$) corresponding to these
contours is very asymmetric with a very steep edge on the $90^\circ$ side.
Only close to the 99\% confidence limit is the timing result consistent with
the scintillation-derived value of $|i-90^\circ|$ of $0^\circ.29 \pm
0^\circ.14$\cite{cmr+05}. We note that the scintillation measurement is based
on the correlation of the scintillation fluctuations of A and B over the short
interval when A is close to superior conjunction (i.e., behind B).  In
contrast, the measurement of $i$ from timing measurements depends on the
detection of significant structure in the post-fit residuals after a portion
of the Shapiro delay is absorbed in the fit for $x_A$ \cite{lcw+01}. 
As shown in Fig.~2,
the Shapiro delay has a signature that is spread over the
whole orbit and hence can be cleanly isolated. 
We also examined the effects
on the Shapiro delay of using only low- or high-frequency data, and found
values of $s$ consistent withing the errors in each case.
The scintillation result is
based on the plasma properties of the interstellar medium and may also be
affected by possible refraction effects in B's magnetosphere. We believe that
the timing result is much less susceptible to systematic errors and is
therefore more secure.

Scintillation observations have also been used to deduce
the system transverse velocity. Ransom et al.\cite{rkr+04} derive a value of
$141\pm 8.5$ km s$^{-1}$ while Coles et al.\cite{cmr+05} obtain $66 \pm 15$ km
s$^{-1}$ after considering the effect of anisotropy in the scattering screen.
Both of these values are in stark contrast to the value of $10 \pm 1$ km
s$^{-1}$ (relative to the Solar system barycentre) obtained from pulsar timing
(Table~1).  We note that the scintillation-based velocity depends on a number
of assumptions about the properties of the effective scattering screen. In
contrast, the proper motion measurement has a clear and unambiguous timing
signature, although the transverse velocity itself scales with the pulsar
distance. Even allowing that unmodelled effects of Earth motion could affect
the published scintillation velocities by about 30 km s$^{-1}$, the
dispersion-based distance would need to be underestimated by a factor of
several to make the velocities consistent.  We believe this is very unlikely,
particularly as the tentative detection of a parallax gives us some confidence
in the dispersion-based distance estimate.  Hence, we believe that
our timing results for both
inclination angle and transverse velocity are less susceptible to systematic
errors and are therefore more secure than those based on scintillation.

We note that, with the inclination angle being significantly different
from $90^\circ$, gravitational lensing effects\cite{rl06} can be
neglected. The implied low space velocity, the comparatively low
derived mass for B and the low orbit eccentricity are all consistent with
the idea that the B pulsar may have formed by a mechanism different to
the usually assumed core-collapse of a helium
star\cite{prps02,ps05}. 
A discussion of its progenitor is presented elsewhere \cite{std+06}.
We also note that, as expected for a
double-neutron-star system, there is no evidence for variation in
dispersion measure as a function of orbital phase.

\paragraph*{Future tests.}

In contrast to all previous tests of GR, we are now reaching the point with
PSR J0737$-$3037A where expressions of PK parameters to only 1PN order may not
be sufficient anymore for a comparison of theoretical predictions with
observations.  In particular, we have measured $\dot\omega$ so precisely
(i.e., to a relative precision approaching $10^{-5}$) that we expect
corrections at the 2PN level\cite{ds88} to be observationally significant
within a few years. These corrections include contributions expected from
spin-orbit coupling\cite{dr74,bo75}.  A future determination of the system
geometry and the measurement of two other PK parameters at a level of
precision similar to that for $\dot\omega$, would allow us to measure the
moment of inertia of a neutron star for the first time\cite{ds88,wex95}.
While this measurement is potentially very difficult, a determination of A's
moment of inertia to a precision of only 30\% would allow us to distinguish
between a large number of proposed equations of state for dense
matter\cite{mbsp04,ls05}. The double pulsar would then not only provide the
best tests of theories of gravity in the strong-field regime as presented here
but would also give insight into the nature of super-dense matter.





\begin{scilastnote}
\item We thank Thibault Damour and Norbert Wex for useful
discussions. The Parkes radio telescope is part of the Australia
Telescope which is funded by the Commonwealth of Australia for
operation as a National Facility managed by CSIRO.  The National Radio
Astronomy Observatory is a facility of the U.S. National Science
Foundation operated under cooperative agreement by Associated
Universities, Inc.  GASP is funded by an NSERC RTI-1 grant to IHS and
by US NSF grants to Donald Backer and David Nice. 
We thank Paul Demorest, Ramachandran and Joeri van
Leeuwen for their contributions to GASP hardward and software
development. IHS holds an NSERC
UFA, and pulsar research at UBC is supported by an NSERC Discovery
Grant. MB, AP and ND'A acknowledge financial support from the Italian
Ministry of University and Research (MIUR) under the national program
{\em Cofin 2003}. FC is supported by NSF, NASA, and NRAO.
\end{scilastnote}

\clearpage
 with an inset
showing an expanded view of the region of principal interest. 

\noindent {\bf Fig. 1.} The tests of general relativity
parameter summarized in a graphical form.  Constraints on the masses of the
two stars (A and B) in the PSR J0737$-$3039A/B binary system. Shaded regions
are forbidden by the individual mass functions of A and B since $\sin i$ must
be $\le 1$.  Other constraining parameters are shown as pairs of lines, where
the separation of the lines indicates the measurement uncertainty. For the
diagonal pair of lines labelled as $R$, representing the mass ratio derived
from the measured semi-major axes of the A and B orbits, the measurement
precision is so good that the line separation only becomes apparent in the
enlarged inset, showing an expanded view of the region of principal interest.
The other constraints shown are based on the measured post-Keplerian (PK)
parameters interpreted within the framework of general relativity. The PK
parameter $\dot\omega$ describes the relativistic precession of the orbit,
$\gamma$ combines gravitational redshift and time dilation, while
$\dot{P}_{\rm b}$ represents the measured decrease in orbital period due to
the emission of gravitational waves.  The two PK parameters $s$ and $r$
reflect the observed Shapiro delay, describing a delay that is added to the
pulse arrival times when propagating through the curved space-time near the
companion. The intersection of all line pairs is consistent with a single
point that corresponds to the masses of A and B. The current uncertainties in
the observed parameters determine the size of this intersection area which is
marked in blue and which reflects the achieved precision of this test of GR
and the mass determination for A and B.
\label{fg:m1m2}

\noindent {\bf Fig. 2.} Measurement of a Shapiro delay demonstrating the 
curvature of space-time.  Timing residuals (differences between observed and
predicted pulse arrival times) are plotted as a function of orbital longitude
and illustrate the Shapiro delay for PSR J0737$-$3039A. (a) Observed timing
residuals after a fit of all model parameters given in Table~1 {\em except}
the Shapiro-delay terms $r$ and $s$ which were set to zero and not included in
the fit. While a portion of the delay is absorbed in an adjustment of the
Keplerian parameters, a strong peak at 90$^\circ$ orbital longitude remains
clearly visible. This is the orbital phase of A's superior conjunction,
i.e.~when it is positioned behind B as viewed from Earth, so that its pulses
experience a delay when moving through the curved space-time near
B. The clear detection of structure in the residuals over the whole orbit
confirms the detection of the Shapiro delay, which is isolated in (b) by
holding all parameters to their best-fit values given in Table~1, except the
Shapiro delay terms which were set to zero.  The line shows the predicted
delay at the centre of the data span. In both cases, residuals were averaged
in $1^\circ$ bins of longitude.

\noindent {\bf Fig. 3.}
Contour plots of the $\chi^2$ distribution in the plane of
the Shapiro-delay parameter $z_s \equiv -\ln(1-s)$ and the mass of the
B pulsar, $m_{\rm{B}}$. The contours correspond to 68\%, 95\% and 99\%
confidence limits.


\clearpage

\noindent Table 1: Parameters for PSR~J0737$-$3039A (A) and PSR~J0737$-$3039B (B).
The values were derived from pulse timing observations using the DD
\cite{dd86} and DDS \cite{ksm+06a} models of the timing
analysis program {\sc tempo} and the Jet Propulsion Laboratory
DE405 planetary ephemeris\cite{sta98c}. Estimated uncertainties, given in
parentheses after the values, refer to the least significant digit of the
tabulated value and are twice the formal 1-$\sigma$ values given by {\sc
  tempo}. The positional parameters are in the DE405 reference frame which is
close to that of the International Celestial Reference System. Pulsar spin
frequencies $\nu \equiv 1/P$ are in barycentric dynamical time (TDB) units at
the timing epoch quoted in Modified Julian Days.  The five Keplerian binary
parameters ($P_b, e, \omega, T_0$, and $x$) are derived for pulsar A. The
first four of these (with an offset of $180^\circ$ added to $\omega$) and the
position parameters were assumed when fitting for B's parameters.  Five
post-Keplerian parameters have now been measured.  An independent fit of
$\dot\omega$ for B yielded a value (shown in square brackets) that is
consistent with the much more precise result for A.  The value derived for A
was adopted in the final analysis (see \cite{som}).  The
dispersion-based distance is based on a model for the interstellar electron
density\cite{cl02}.

\begin{table*}[ht]
\begin{center}
\begin{scriptsize}
\begin{tabular}{lcc}
\hline
Timing parameter & PSR~J0737$-$3039A & PSR~J0737$-$3039B \\ \hline
Right Ascension $\alpha$  & $07^{\rm{h}}37^{\rm{m}}51^{\rm{s}}.24927(3)$ & $-$ \\
Declination $\delta$  & $-30^\circ 39' 40''.7195(5)$ & $-$ \\
Proper motion in the RA direction (mas yr$^{-1}$) & $-3.3(4)$ & $-$ \\
Proper motion in Declination (mas yr$^{-1}$) & $2.6(5)$ & $-$ \\
Parallax, $\pi$ (mas) & 3(2) & $-$ \\
Spin frequency $\nu$ (Hz) & 44.054069392744(2) &  0.36056035506(1) \\
Spin frequency derivative $\dot\nu$ (s$^{-2}$) & $-3.4156(1) \times 10^{-15}$ &
$-0.116(1)\times 10^{-15}$ \\
Timing Epoch (MJD) & 53156.0 & 53156.0 \\
Dispersion measure DM (cm$^{-3}$pc) & 48.920(5) &  $-$\\
Orbital period $P_b$ (day) & 0.10225156248(5) & $-$ \\
Eccentricity $e$ & 0.0877775(9) & $-$ \\
Projected semi-major axis $x=(a/c)\sin i$ (s) & 1.415032(1) & 
1.5161(16) \\
Longitude of periastron $\omega$ (deg) & 87.0331(8) & 87.0331 + 180.0 \\
Epoch of periastron $T_0$ (MJD) & 53155.9074280(2) & $-$ \\
Advance of periastron $\dot\omega$ (deg/yr) & 16.89947(68) & [16.96(5)] \\
Gravitational redshift parameter $\gamma$ (ms) & 0.3856(26) & $-$ \\
Shapiro delay parameter $s$ & $0.99974(-39,+16)$ & $-$ \\
Shapiro delay parameter $r$ ($\mu$s) & 6.21(33) & $-$ \\
Orbital period derivative $\dot P_b$ & $-1.252(17)\times 10^{-12}$ &$-$ \\ \hline
Timing data span (MJD) & 52760 -- 53736 & 52760 -- 53736 \\
Number of time offsets fitted & 10 & 12 \\ 
RMS timing residual $\sigma$ ($\mu$sec) & 54 & 2169 \\
Total proper motion (mas yr$^{-1}$) & \multicolumn{2}{c}{4.2(4)} \\ 
Distance $d$(DM) (pc) & \multicolumn{2}{c}{$\sim 500$} \\ 
Distance $d(\pi)$ (pc) & \multicolumn{2}{c}{$200-1000$} \\ 
Transverse velocity ($d=500$ pc) (km s$^{-1}$) & \multicolumn{2}{c}{10(1)} \\ 
Orbital inclination angle (deg) & \multicolumn{2}{c}{88.69(-76,+50)} \\ 
Mass function ($M_\odot$) &  $0.29096571(87)$ & $0.3579(11)$ \\ 
Mass ratio, $R$ &   \multicolumn{2}{c}{1.0714(11)}  \\ 
Total system mass ($M_\odot$) &   \multicolumn{2}{c}{2.58708(16)}     \\ 
Neutron star mass ($m_\odot$) & 1.3381(7) & 1.2489(7) \\
\hline
\end{tabular}
\end{scriptsize}
\end{center}
\label{tab:params}
\end{table*}

\clearpage
\noindent
Table 2: Four independent tests of GR provided by the double pulsar.
The second column lists the observed PK parameters obtained by fitting
a DDS timing model to the data.  The third column lists the values
expected from general relativity given the masses determined from the
intersection point of the mass ratio $R$ and the periastron advance
$\dot{\omega}$. The last column gives the ratio of the observed to
expected value for each test. Uncertainties refer to the last quoted
digit and were determined using Monte Carlo methods.

\begin{table*}[ht]
\begin{scriptsize}
\begin{center}
\begin{tabular}{cccc}
PK parameter & Observed & GR expectation & Ratio \\
\hline
$\dot{P}_{\rm b}$ & 1.252(17) & 1.24787(13) & 1.003(14) \\
$\gamma$ (ms) & 0.3856(26) & 0.38418(22)& 1.0036(68)    \\
$s$ & 0.99974($-39$,+16) & 0.99987($-48$,+13) & 0.99987(50) \\
$r(\mu$s) & 6.21(33) & 6.153(26) & 1.009(55)  \\
\hline
\end{tabular}
\end{center}
\end{scriptsize}
\label{tab:compare}
\end{table*}

\newpage

\centerline{\psfig{file=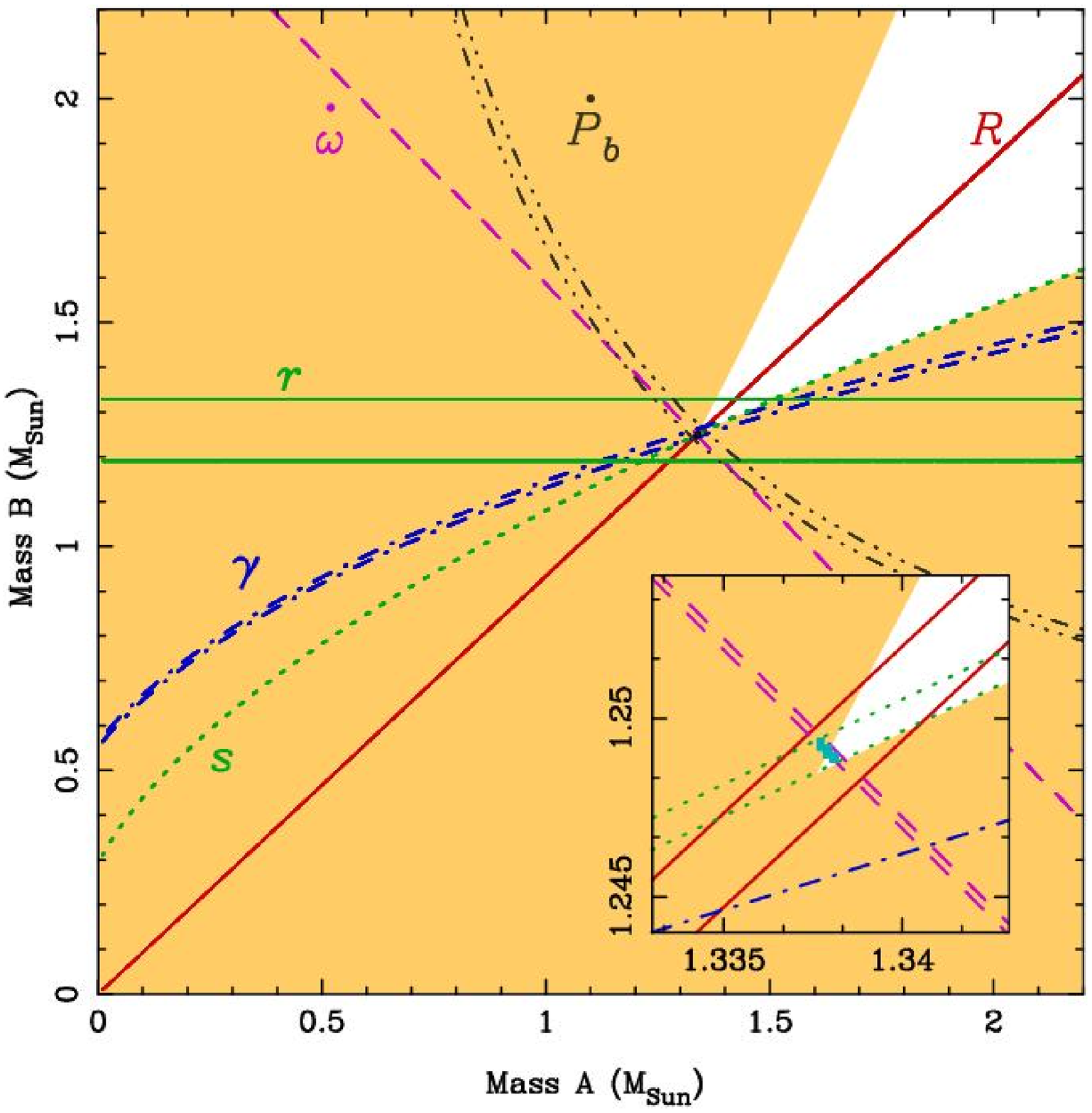,angle=-90,width=14cm}\\ Fig.~1}

\newpage
\centerline{\psfig{file=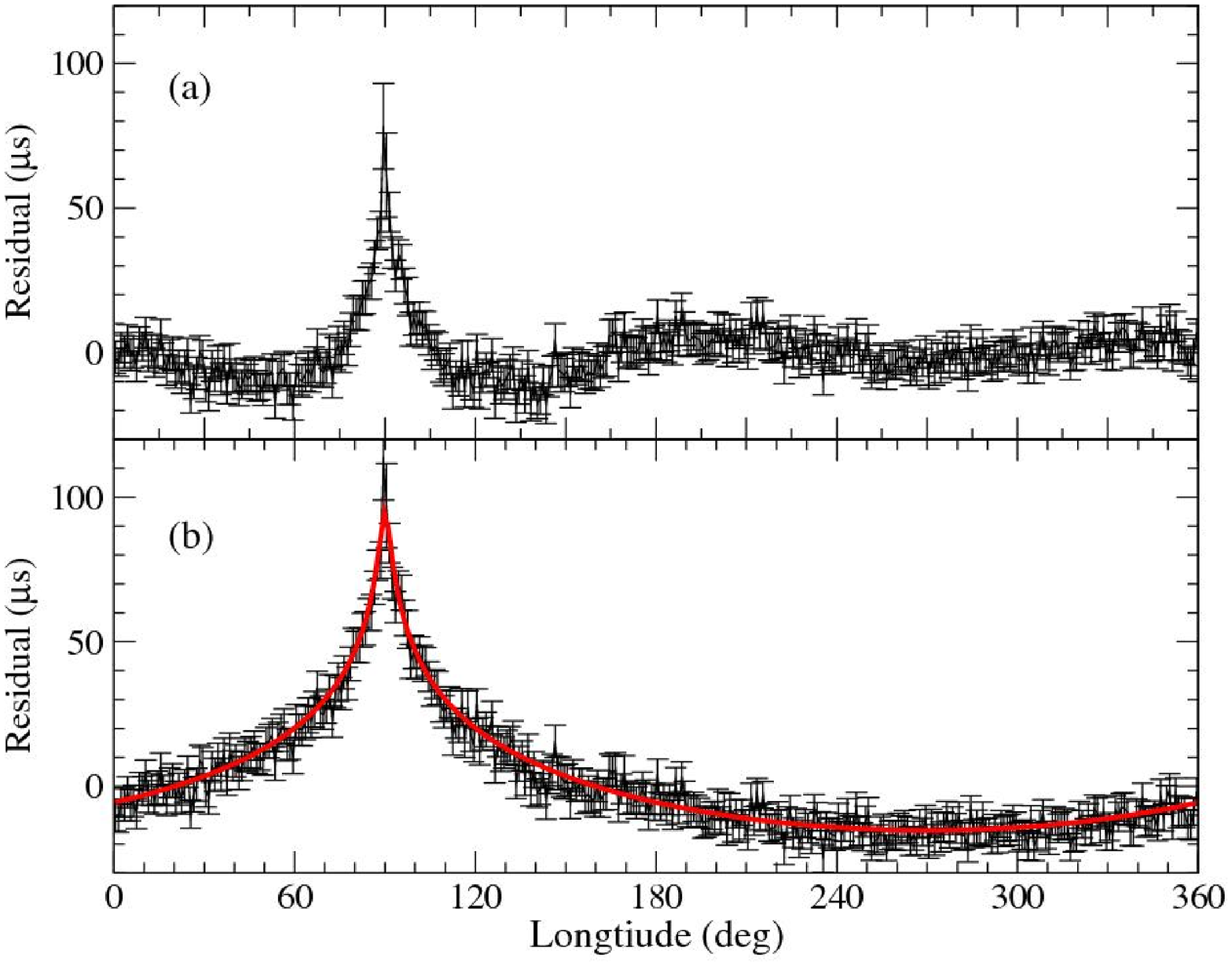,angle=-90,width=14cm}\\ Fig.~2}

\newpage
\centerline{\psfig{file=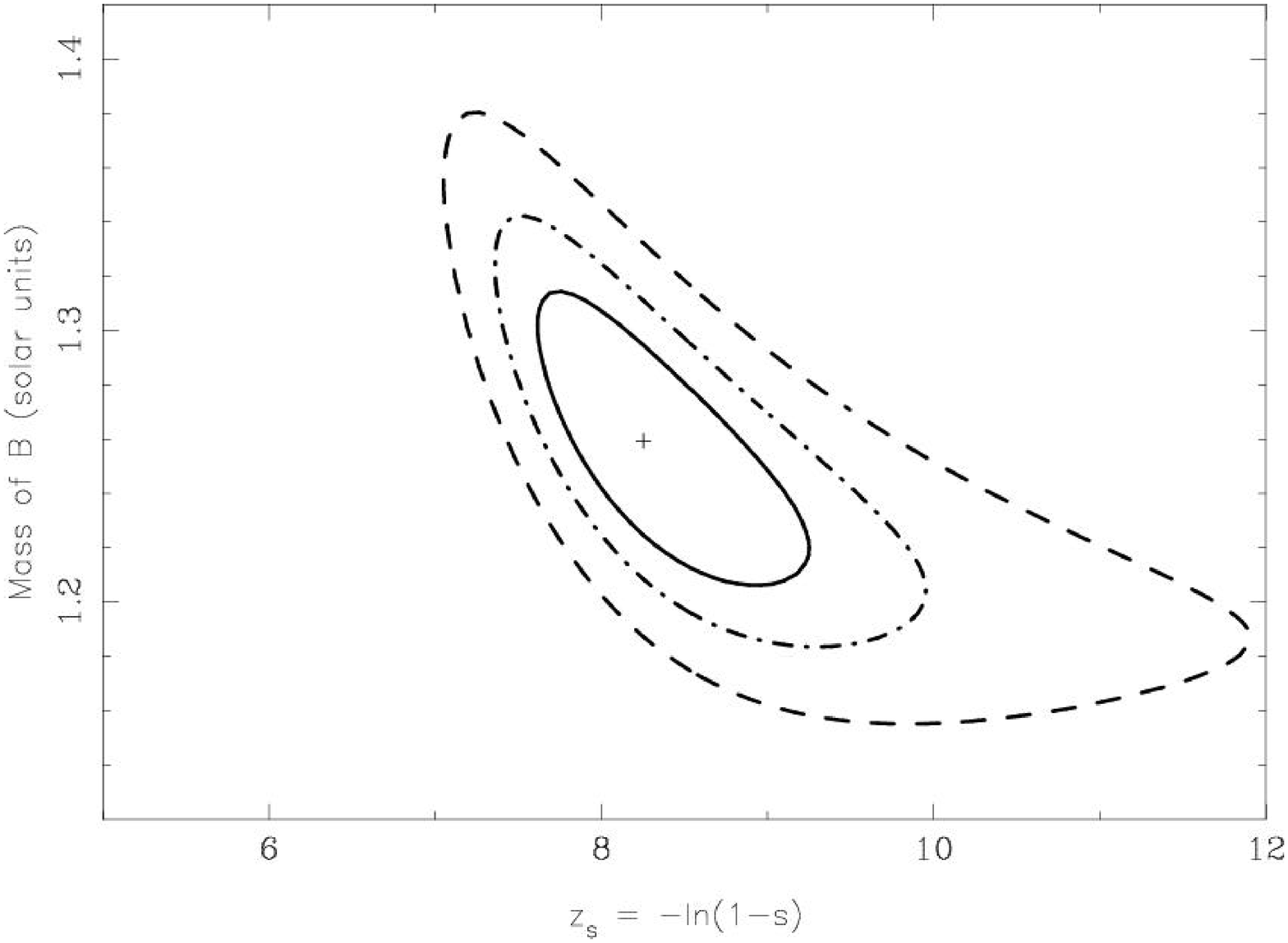,angle=-90,width=15cm}\\Fig.~3}

\newpage

\centerline{\large \bf Supporting Online Material}

\section{Observing systems}

The experimental data presented in the main paper are based on pulsar
timing observations at several frequencies between 320 MHz and 3100
MHz using the Parkes radio telescope in Australia, the Lovell
radio telescope at Jodrell Bank Observatory, UK, and the Green Bank
Telescope (GBT) in the USA, between 2003 April and 2006
January. Details of the observing systems are summarized in
Supporting Table~1.

At the Parkes 64-m radio telescope observations were carried out using the
centre beam of the 20-cm multibeam receiver and a coaxial 10cm/50cm receiver.
For each of these cryogenically cooled receivers, two orthogonally polarized
signals were amplified and down-converted to an intermediate frequency. These
signals were transferred to band splitters and fed into a filterbank system
(FB) for each polarization of each feed. The output of each filter was
detected and summed with its corresponding polarization pair. These summed
outputs were high-pass filtered and integrated for the sampling interval of 80
$\mu$s and then one-bit digitised.  While the original frequency channels were
folded with a reference frequency corresponding to the band centre, timing was
performed on sub-bands.

Observations at the GBT utilized two different data acquisition systems.  The
Berkeley-Caltech Pulsar Machine (BCPM) is a flexible filterbank
system\cite{bdz+97}, with which we collected 4-bit summed-polarization data.
The Green Bank Astronomical Signal Processor (GASP) carries out 8-bit
Nyquist-sampling of the incoming dual-polarization signal, after which it
performs coherent dedispersion in software on a Linux-based cluster for each
of several 4-MHz channels\cite{drb+04a,fsb+04}. The data stream is then
detected, and the two polarizations are usually flux-calibrated before
summation using a diode noise source as a reference.

At Jodrell Bank we used a incoherently dedispersing filterbank system.  Its
parameters are summarized in Table 1, while details of the observing system
can be found in ref.~4\nocite{gl98}.

\section{Dedispersion}

Since the interstellar medium (ISM) is ionized, 
the propagation speed of radio pulses
depends on their radio frequency with pulses emitted at a high radio frequencies
arriving earlier than low-frequency pulses. Unless this effect is accounted for,
pulses will be broadened over the finite observing bandwidth. Two
dedispersion techniques are in use. For ``incoherent dedispersion'', the
bandwidth is sub-divided into a number of frequency channels which are
detected and sampled independently.  Dispersion smearing is thereby reduced to
the smearing across an individual filterbank channel. The ``coherent
dedispersion'' technique involves the application of an inverse 
``ISM-filter'' to the raw
voltage data received from the antenna \cite{hr75}. This technique is
computationally more intensive but removes the effects of dispersion
completely. 

At Parkes and Jodrell Bank we obtained incoherently dedispersed data using the
filterbank systems listed in Table 1. The resulting profiles were summed
across frequency channels with appropriate delays to remove the effects of
interstellar dispersion. For the wide-bandwidth Parkes data, where the
original frequency channels were folded with a reference frequency
corresponding to the band centre, timing was performed on sub-bands. The
number of sub-bands was chosen such that the dispersion delay across the
sub-bands was significantly smaller than the overall timing precision.
Analysis of TOA data separately for the different sub-bands properly accounts
for the fact that data at different frequencies received at a given time
correspond to different orbital phases at emission due to the differential
dispersion delay (see e.g.~\cite{hem06}).

At the GBT, the BCPM data were divided in four frequency sub-bands, separately
dedispersed, folded and timed. In contrast, each GASP 4-MHz channel was
coherently dedispersed and folded using the channel centre frequency as a
reference. The GASP channels were then summed appropriately to give a single
TOA for each integration.

\section{Pulse Time-of-Arrival analysis}

Pulse times-of-arrival (TOAs) were computed by
correlating the observed pulse profiles with synthetic noise-free templates
(see Figs.~1 and 2; cf.~ref.~\cite{kll+99}).
All datasets obtained at different
epochs and frequencies with different data acquisition hardware and
telescopes were studied for possible systematic errors and artificial
correlations.  Firstly, correlations between successive TOAs were
investigated by computing the post-fit root-mean-square (rms) timing
residuals with averaging of consecutive TOAs, expecting that the rms
residual should decrease with the square-root of the number of
averaged TOAs.  Datasets with significant deviations from this
expected scaling were excluded from the analysis. Secondly, for the
GBT observations where we recorded data with two different data
acquisition systems in parallel, we preferred to use to more accurate
GASP data and only used BCPM data if no GASP TOAs were available
within 2 minutes of a BCPM TOA.  Thirdly, the uncertainties of the
TOAs in the remaining datasets were studied by inspecting the reduced
$\chi^2$ achieved in the fit of the timing model.  For most datasets
we applied a small quadrature addition and a scaling factor to the
uncertainties to obtain the expected value of $\chi^2_{\rm red}=1$. No
adjustments to the TOA uncertainties were needed for the GASP data;
this is not surprising as the 8-bit sampling provides excellent
profile fidelity.  Finally, all retained datasets were combined in a
weighted least-squares fit of the DD and DDS models. Following these
fits, we verified that the $\chi^2_{\rm red}$ for each data subset was
still close to unity.  A total of 131,416 arrival times were included
in the final analysis of A while 507 TOAs were used for B, most at
frequencies close to 820 MHz and 1400 MHz. The much smaller number of
TOAs for B results from several factors: JBO data were not used, the
integration time for B was a factor of ten larger than for A, the data were
summed over the entire observed frequency band, only about 20\% of the
orbit was used and finally, even in the analysed regions, B was often
too weak to give a significant TOA.  Figures~3 and 4
summarise the TOA distributions for the different
observatories for pulsars A and B respectively. Finally, we present
the covariance matrix as computed by TEMPO for the fit of the DDS
timing model in Table 2.


\begin{thebibliography}{10}

\bibitem{wil01}
C.~{Will}, {\it Living Reviews in Relativity\/} {\bf 4}, 4 (2001).

\bibitem{bit03}
B.~Bertotti, L.~Iess, P.~Tortora, {\it Nature\/} {\bf 425}, 374 (2003).

\bibitem{wtb04}
J.~G. Williams, S.~G. Turyshev, D.~H. Boggs, {\it Phys. Rev. Lett.\/} {\bf 93},
  261101 (2004).

\bibitem{de98}
T.~Damour, G.~{Esposito-Far{\` e}se}, {\it Phys. Rev. D\/} {\bf 58}, 1 (1998).

\bibitem{tw89}
J.~H. Taylor, J.~M. Weisberg, {\it ApJ\/} {\bf 345}, 434 (1989).

\bibitem{sttw02}
I.~H. Stairs, S.~E. Thorsett, J.~H. Taylor, A.~Wolszczan, {\it ApJ\/} {\bf
  581}, 501 (2002).

\bibitem{bdp+03}
M.~{Burgay}, {\it et~al.\/}, {\it Nature\/} {\bf 426}, 531 (2003).

\bibitem{lbk+04}
A.~G. Lyne, {\it et~al.\/}, {\it Science\/} {\bf 303}, 1153 (2004).

\bibitem{bpm+05}
M.~{Burgay}, {\it et~al.\/}, {\it ApJ\/} {\bf 624}, L113 (2005).

\bibitem{dd86}
T.~Damour, N.~Deruelle, {\it Ann. Inst. H. Poincar\'e (Physique Th\'eorique)\/}
  {\bf 44}, 263 (1986).

\bibitem{dt92}
T.~Damour, J.~H. Taylor, {\it Phys. Rev. D\/} {\bf 45}, 1840 (1992).

\bibitem{ds88}
T.~Damour, G.~Sch{\"a}fer, {\it Nuovo Cim.\/} {\bf 101}, 127 (1988).

\bibitem{som} Supporting Online Material

\bibitem{kll+99}
M.~Kramer, {\it et~al.\/}, {\it ApJ\/} {\bf 526}, 957 (1999).

\bibitem{mkp+05}
R.~N. {Manchester}, {\it et~al.\/}, {\it ApJ\/} {\bf 621}, L49 (2005).

\bibitem{tempo} {\it http://www.atnf.csiro.au/research/pulsar/tempo}.

\bibitem{dd85}
T.~Damour, N.~Deruelle, {\it Ann. Inst. H. Poincar\'e (Physique Th\'eorique)\/}
  {\bf 43}, 107 (1985).

\bibitem{ksm+06a}
M.~Kramer, {\it et~al.\/}, {\it {Annalen der Physik}\/} {\bf 15}, 34 (2006).

\bibitem{wt05}
J.~M. {Weisberg}, J.~H. {Taylor}, {\it {Binary Radio Pulsars}\/}, F.~Rasio,
  I.~H. Stairs, eds. (Astronomical Society of the Pacific, San Francisco,
  2005), pp. 25--31.

\bibitem{twdw92}
J.~H. Taylor, A.~Wolszczan, T.~Damour, J.~M. Weisberg, {\it Nature\/} {\bf
  355}, 132 (1992).

\bibitem{lyu05}
M.~{Lyutikov}, {\it MNRAS\/} {\bf 362}, 1078 (2005).

\bibitem{dt91}
T.~Damour, J.~H. Taylor, {\it ApJ\/} {\bf 366}, 501 (1991).

\bibitem{cl02}
J.~M. {Cordes}, T.~J.~W. {Lazio}, 
{\it NE2001. I. A New Model for the Galactic Distribution
  of Free Electrons and its Fluctuations} (2002). {astro-ph/0207156}.

\bibitem{kg89}
K.~Kuijken, G.~Gilmore, {\it MNRAS\/} {\bf 239}, 571 (1989).

\bibitem{de06}
T.~Damour, G.~Esposito-Far`ese, to appear. (2006).

\bibitem{cmr+05}
W.~A. Coles, M.~A. McLaughlin, B.~J. Rickett, A.~G. Lyne, N.~D.~R. Bhat, {\it
  ApJ\/} {\bf 623}, 392 (2005).

\bibitem{lcw+01}
C.~Lange, {\it et~al.\/}, {\it MNRAS\/} {\bf 326}, 274 (2001).

\bibitem{rkr+04}
S.~M. Ransom, {\it et~al.\/}, {\it ApJ\/} {\bf 609}, L71 (2004).

\bibitem{rl06}
R.~R. {Rafikov}, D.~{Lai}, {\it Phys. Rev. D\/} {\bf 73}, 063003 (2006).

\bibitem{prps02}
E.~{Pfahl}, S.~{Rappaport}, P.~{Podsiadlowski}, H.~{Spruit}, {\it ApJ\/} {\bf
  574}, 364 (2002).

\bibitem{ps05}
T.~{Piran}, N.~J. {Shaviv}, {\it Phys. Rev. Lett.\/} {\bf 94}, 051102 (2005).

\bibitem{std+06}
I.~H. Stairs, S.~E. Thorsett, R.~J. Dewey, M.~Kramer, C.~McPhee, {\it MNRAS\/}
  in press (2006).

\bibitem{dr74}
T.~Damour, R.~Ruffini, {\it Academie des Sciences Paris Comptes Rendus
  Ser.\,Scie.\,Math.\/} {\bf 279}, 971 (1974).

\bibitem{bo75}
B.~M. Barker, R.~F. O'Connell, {\it ApJ\/} {\bf 199}, L25 (1975).

\bibitem{wex95}
N.~Wex, {\it Class. Quantum Grav.\/} {\bf 12}, 983 (1995).

\bibitem{mbsp04}
I.~A. {Morrison}, T.~W. {Baumgarte}, S.~L. {Shapiro}, V.~R. {Pandharipande},
  {\it ApJ\/} {\bf 617}, L135 (2004).

\bibitem{ls05}
J.~M. {Lattimer}, B.~F. {Schutz}, {\it ApJ\/} {\bf 629}, 979 (2005).


\bibitem{sta98c}
E.~M. {Standish}, {\it A\&A\/} {\bf 336}, 381 (1998).

\end{thebibliography}

\begin{thebibliography}{1}

\bibitem{bdz+97}
D.~C. Backer, {\it et~al.\/}, {\it PASP\/} {\bf 109}, 61 (1997).

\bibitem{drb+04a}
P.~{Demorest}, {\it et~al.\/}, {\it American Astronomical Society Meeting
  Abstracts\/} {\bf 205},  (2004).

\bibitem{fsb+04}
R.~D. {Ferdman}, {\it et~al.\/}, {\it American Astronomical Society Meeting
  Abstracts\/} {\bf 205},  (2004).

\bibitem{gl98}
D.~M. Gould, A.~G. Lyne, {\it MNRAS\/} {\bf 301}, 235 (1998).

\bibitem{hr75}
T.~H. {Hankins}, B.~J. {Rickett}, {\it Methods in Computational Physics Volume
  14 --- Radio Astronomy\/} (Academic Press, New York, 1975), pp. 55--129.

\bibitem{hem06}
G.~B. {Hobbs}, R.~T. {Edwards}, R.~N. {Manchester}, {\it MNRAS\/} {\bf 369},
  655 (2006).

\bibitem{kll+99}
M.~Kramer, {\it et~al.\/}, {\it ApJ\/} {\bf 526}, 957 (1999).

\end{thebibliography}

\clearpage

\noindent {\bf Supporting Table 1.}
Summary of the observing systems used for timing
  observations of the double pulsar.

\begin{table*}[ht]
\begin{center}
\begin{scriptsize}
\begin{tabular}{l|lcccccc}
\hline\hline
Telescope     &  Instrument  & Centre  &  Gain & $T_{sys}$  & Sample    & Bandwidth & Number   \\
              &   & freq. (MHz)    &  (K/Jy) & (K)   & interval ($\mu$s) & (MHz)  & of channels  \\
\hline\hline                                                                                                                                              
              &            &  680               &   0.66 &  45  &  80   &  64  &  128      \\
 Parkes       &  FB        &  1374               &  0.74 &  22  &  80   &  256 &  512      \\
              &            &  3030               &  0.62 &  28  &  80   &  768 &  256      \\
\hline                                                                                                                                                            
  GBT         & BCPM         &  820              &  2.0  &  25  &  72   &  48   &  96       \\
              &              &  1400             &  2.0  &  20  &  72   &  96   &  96       \\
\cline{2-8}                                                                                                                                               
              & GASP         &  340              &  2.0  &  70  &  0.25 &  16   &  4        \\
              &              &  820              &  2.0  &  25  &  0.25 &  64   &  16$^a$   \\
              &              &  1400             &  2.0  &  20  &  0.25 &  64   &  16$^a$   \\
\hline                                                                                                                                                            
 Jodrell Bank &  FB          &  610              &  1.1  &  32  &  44.4 &  8    &   32       \\
              &  FB          &  1396             &  1.1  &  32  &  44.4 &  64    &   64   \\
\hline
\end{tabular}
\vspace{20pt}
\newline
$^a$The number of channels and hence bandwidth that was used varied
occasionally within a given session due to the removal of 
channels contaminated with radio frequency interference and/or
occasional recording disk-space limitations.
\end{scriptsize}
\end{center}
\end{table*}

\clearpage
\bigskip
\noindent {\bf Supporting Table 2.}
Covariance matrix as computed by TEMPO for a fit of the DDS timing
model to the TOAs of A.

\begin{table*}[ht]
\begin{center}
\tiny
\begin{tabular}{r|rrrrrrrrrrrrrrrrrr}
       &     $\nu$ & $\dot{\nu}$  & Dec  &  RA & PMDec & PMRA & $x$ & $e$ &
       $T_0$ & $P_{\rm b}$ & $\omega$ & $\dot\omega$ & $\gamma$ &  DM & $\pi$
       & $\dot{P}_{\rm b}$ & $z_s$ & $m_2$ \\  
\hline
   $\nu$&    1.00 & & & & & & & & & & & & & & & & & \\ 
   $\dot\nu$&  -0.76 & 1.00 & & & & & & & & & & & & & & & & \\
    Dec&   0.16 & -0.31 &   1.00 & & & & & & & & & & & & & & & \\
     RA&   0.10 & -0.08 & 0.18 & 1.00 & & & & & & & & & & & & & & \\
   PMDec&  -0.25 &  0.39 & -0.83 & -0.16 &  1.00 & & & & & & & & & & & & & \\
   PMRA&   0.01 & -0.28&  0.04& -0.71&  0.12&  1.00 & & & & & & & & & & & & \\
   $x$ &  -0.02 & 0.02 & 0.01&  0.00& -0.01& -0.01&  1.00& & & & & & & & & & & \\
   $e$&   0.00 & 0.00 & 0.01& -0.01& -0.02& -0.01&  0.66&  1.00& & & & & & & &
       & & \\
   $T_0$&  -0.54 & 0.43 &-0.01 & 0.02 & 0.02& -0.03&  0.00&  0.01&  1.00 & & & &
       & & & & & \\
   $P_{\rm b}$&   0.47&  -0.47&  0.01 &-0.02& -0.02&  0.02& -0.15& -0.15& -0.85&
       1.00& & & & & & & & \\
  $\omega$&  -0.54 & 0.42& -0.01 & 0.02 & 0.02& -0.03 & 0.06 & 0.01&  0.99& -0.84
       & 1.00 & & & & & & & \\
  $\dot\omega$&   0.47& -0.48 & 0.01 &-0.02& -0.02&  0.02& -0.15 &-0.15& -0.85 & 1.00&
       -0.84&  1.00& & & & & & \\
  $\gamma$&  -0.02& -0.02&  0.00&  0.01 & 0.00&  0.00&  0.44&  0.01& -0.03&  0.02
       & 0.10 & 0.03&  1.00& & & & & \\
     DM&  -0.02 & 0.02& -0.01&  0.01&  0.03 & 0.02&  0.00&  0.00&  0.00&  0.00&
       0.00&  0.00&  0.00&  1.00& & & & \\
   $\pi$&   0.05& -0.02& -0.21&  0.16&  0.20& -0.12& -0.03& -0.04& -0.01& -0.01&
       0.00& -0.01&  0.02& -0.02&  1.00& & & \\
 $\dot{P}_{\rm b}$&   0.01& -0.02 & 0.00 & 0.03 & 0.00& -0.02&  0.05 & 0.00& -0.03& -0.06&
       0.01&  0.01&  0.13&  0.00& -0.01&  1.00& & \\
 $z_s$&  -0.01 & 0.02 & 0.01& -0.01 &-0.01&  0.00 & 0.55&  0.44 & 0.01& -0.10&
       0.01& -0.10& -0.01&  0.00& -0.01&  0.00 & 1.00&   \\
   $m_2$&   0.01& -0.03& -0.01&  0.00&  0.01&  0.01& -0.85& -0.75& -0.02&  0.18& -0.02&  0.18& -0.02 & 0.00&  0.03&  0.00& -0.72&  1.00\\
\hline
\end{tabular}
\end{center}
\end{table*}

\clearpage

\bigskip
\noindent {\bf Supporting Figure 1.}
Pulse profile templates used for TOA determinations for pulsar A.

\bigskip
\noindent {\bf Supporting Figure 2.}
Regions of orbital phase (hatched) used for timing of pulsar B and
  pulse profile templates for these phases derived from and used for
  the 820 MHz GBT observations in May 2005. Each of the template plots
  covers a range of $60/360=0.17$ in pulse phase. Similar but different
  templates were used for other frequencies and epochs.  While B was
  clearly detectable in these three regions, it is actually brightest
  in the two cross-hatched regions, but because the shape of the
  profile evolves quickly and dramatically in these regions, they were
  excluded from the timing analysis.

\noindent {\bf Supporting Figure 3.}
Timing residuals obtained for pulsar A 
for the three telescopes and their
distribution. The upper panel shows the distribution of
observations in frequency.

\bigskip
\noindent {\bf Supporting Figure 4.}
Timing residuals obtained for pulsar B
for Parkes and the GBT and their
distribution. The upper panel shows the distribution of
observations in frequency.

\clearpage
\centerline{\psfig{file=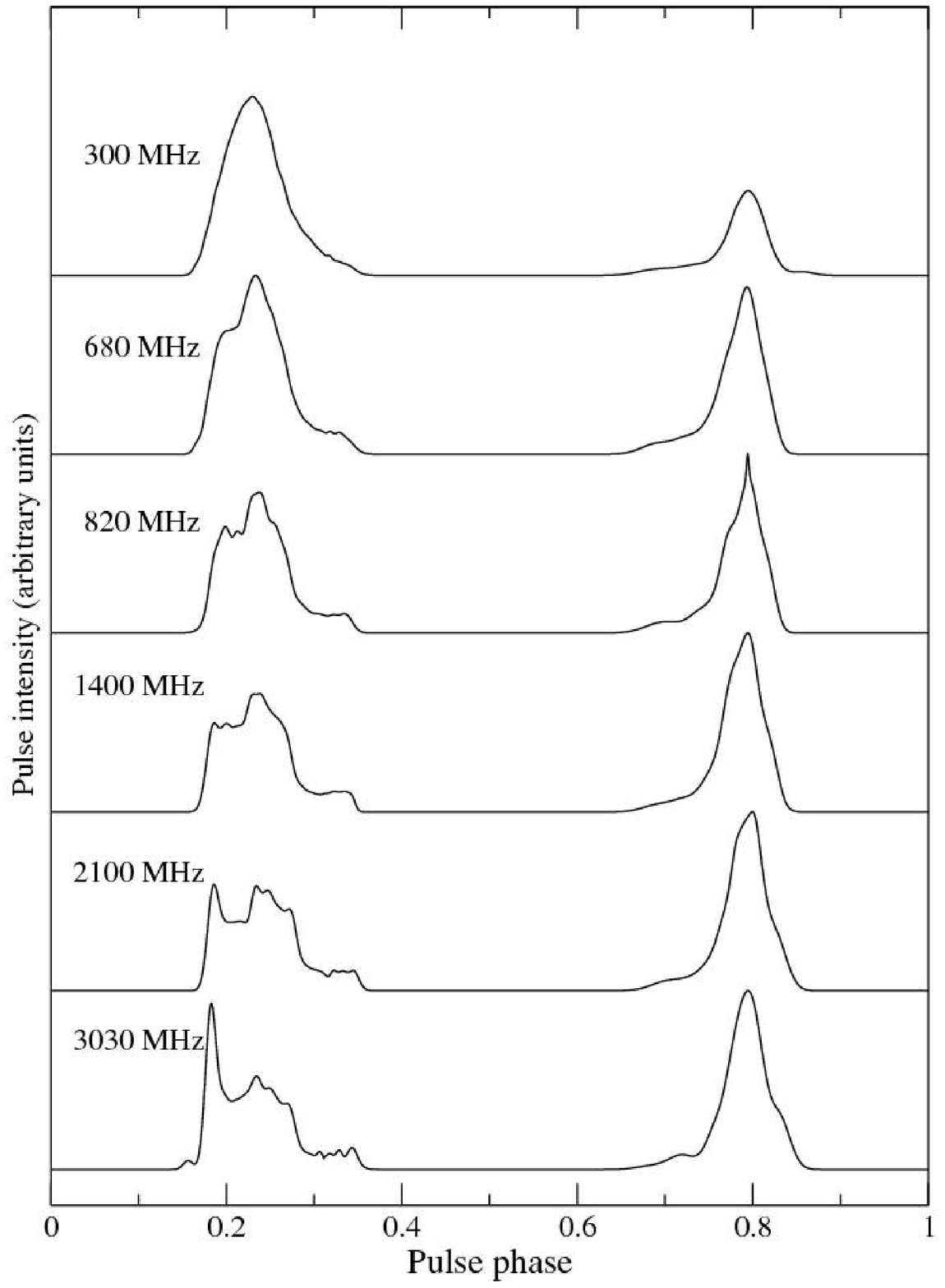,width=15cm}\\Fig.~1}

\clearpage
\centerline{\psfig{file=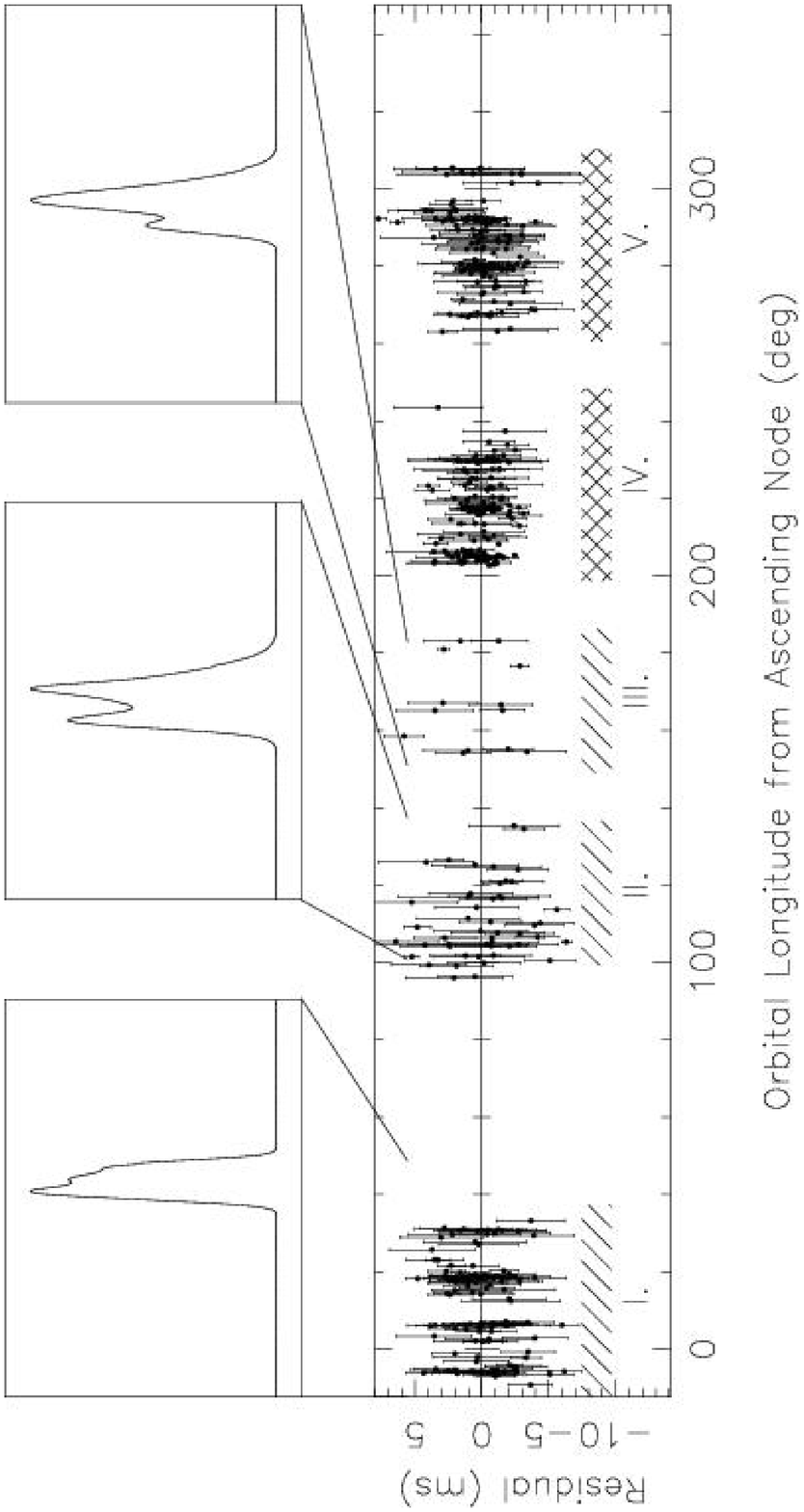,angle=-90,width=15cm}\\Fig.~2}

\clearpage
\centerline{\psfig{file=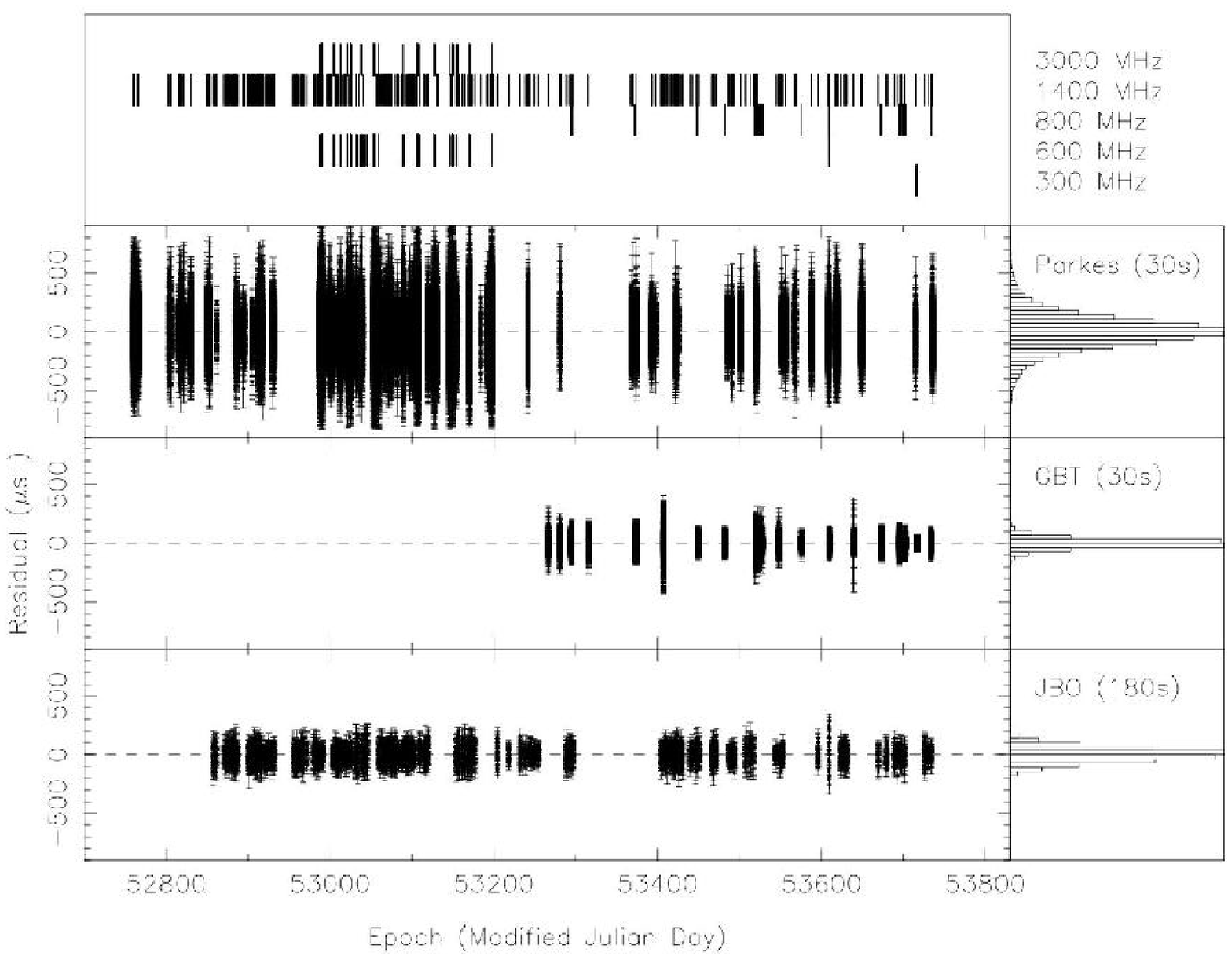,angle=-90,width=15cm}\\Fig.~3}

\clearpage
\centerline{\psfig{file=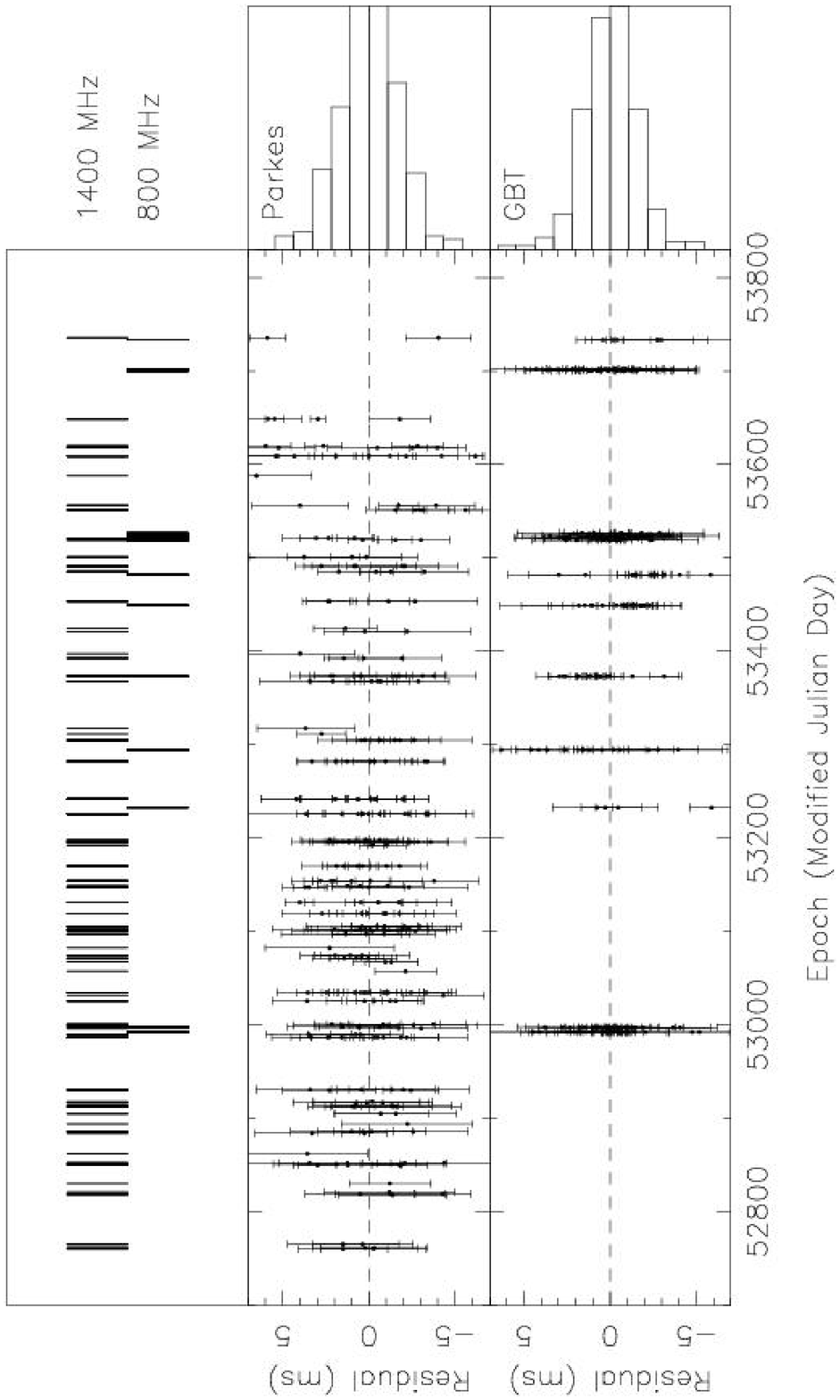,angle=-90,width=15cm}\\Fig.~4}

\end{document}